\documentclass[twocolumn,prl,showpacs,amsmath,amssymb,nofootinbib,superscriptaddress,preprintnumbers]{revtex4}

\usepackage{dcolumn}
\usepackage{amsmath}
\usepackage{graphicx}
\usepackage{rotating}
\usepackage{amsfonts}
\usepackage{amssymb}
\usepackage{subfigure}

\usepackage{graphicx}
\usepackage{epsfig}
\newcommand{\JJ}[1]{#1}

\begin{document}

\preprint{IPPP/06/60}
\preprint{DCPT/06/120}
\preprint{DESY 06-133}

\title {
Illuminating the Hidden Sector of String Theory\\ by
Shining Light through a Magnetic Field}

\author{Steven A. Abel}
\affiliation{Centre for Particle Theory, Durham University, Durham, DH1 3LE, UK}
\author{Joerg Jaeckel}
\affiliation{Deutsches Elektronen-Synchrotron DESY, Notkestra\ss e 85, D-22607 Hamburg, Germany}
\author{Valentin V. Khoze}
\affiliation{Centre for Particle Theory, Durham University, Durham, DH1 3LE, UK}
\author{Andreas Ringwald}
\affiliation{Deutsches Elektronen-Synchrotron DESY, Notkestra\ss e 85, D-22607 Hamburg, Germany}

\begin{abstract}
Many models of physics beyond the Standard Model predict minicharged particles
to which current and near future low-energy experiments are highly sensitive.
Such minicharges arise generically from kinetic-mixing in
theories containing at least two $U(1)$ gauge factors. Here,  we point
out  that the required multiple $U(1)$ factors, the size of
kinetic-mixing, and suitable matter representations to allow for a detection in the near future
occur naturally in the context of string theory embeddings of the Standard Model.
A detection of minicharged particles in
a low energy experiment would likely be a signal of an underlying string theory
and may provide a means of testing it.
\end{abstract}

\pacs{11.25.-w, 11.25.Wx, 12.20.-m, 14.80.-j}

\maketitle

The absorption probability and the propagation speed of polarized
light propagating in a magnetic field may depend on the
relative orientation of the polarization and the magnetic field.
These effects are known as vacuum magnetic dichroism and birefringence,
respectively.

In 2006, the PVLAS
collaboration~\cite{Zavattini:2005tm} reported an anomalously large rotation of the
polarization
plane of light after its passage through a transverse magnetic field in vacuum.
In Ref.~\cite{Gies:2006ca}, it was shown that such a signal may be originating from the
dichroism caused by pair production of minicharged fermions of sub-eV mass and fractional electric charge.
More recent measurements by the PVLAS collaboration with an improved apparatus \cite{Zavattini:2007ee}
did not confirm this signal.
Accordingly these new measurements provide a bound of roughly~\cite{Ahlers:2006iz,Ahlers:2007qf,Ahlers:2007rd}
\begin{equation}
\label{pvlas_mcf_eps}
\epsilon\equiv Q_f/e \lesssim {\rm{few}}\times 10^{-7},\quad {\rm{for}} \quad m_f\lesssim 0.1 \,{\rm{eV}}.
\end{equation}
This is the best known laboratory bound on the existence of light minicharged particles demonstrating
that optical experiments are a powerful tool to
search for such particles.

Moreover, motivated by the initial PVLAS result it has been demonstrated that minicharged particles and
hidden-sector $U(1)$
gauge bosons can also be searched for in a variety of other low-energy
laboratory experiments and significant improvements in the sensitivity are expected in the near future
(see discussion at the end of the paper).

In this letter, we argue that
models with minicharged fermions can naturally
and generically arise in string theory.
Detection of minicharged particles would
therefore not only address
the fundamental question of charge quantization,
but also provide insight into the underlying theory of nature.

Particles with a small, unquantized charge arise very naturally in so-called
paraphoton~\cite{Okun:1982xi} models, containing, beyond
the usual electromagnetic $U(1)$ gauge factor, at least one additional hidden-sector
$U(1)$ factor. The basic observation is that particles with paracharge get
an induced electric charge proportional to some small mixing angle
between the kinetic terms of photons and paraphotons~\cite{Holdom:1985ag}.
Moreover, in models containing more than one paraphoton
with at least one paraphoton being exactly massless and one light,
keV~$\gg m_{{\gamma^\prime }}\neq 0$,
the prohibitively strong astrophysical bounds on the fractional charge,
$\epsilon\lesssim\, 2\times 10^{-14}$,
for $m_f\,\lesssim$\,few keV, arising from energy loss considerations
of stars~\cite{Davidson:2000hf}, can be
relaxed considerably. In a simple model analysed in~\cite{Masso:2006gc},
there are two paraphotons: one massless and one light, and
the fermion transforms in the bifundamental representation of these two
$U(1)$ factors.
In vacuum, the fermion acquires an electric charge $\epsilon$ due to a
kinetic-mixing between the photon and the two paraphotons. Importantly, however,
this electric charge
is reduced in the stellar plasma by a multiplicative
factor $m_{{\gamma^\prime}}^2/\omega_p^2$, where $\omega_p\sim$~few keV is the
plasma frequency.
This charge screening mechanism
is caused by a partial cancellation between two paraphotons
interacting with the
bifundamental fermion~\cite{Masso:2006gc}.
The vacuum value~(\ref{pvlas_mcf_eps}) is therefore perfectly compatible with
astrophysical bounds (as well as cosmological bounds based on big bang nucleosynthesis) as long as
\begin{equation}
\label{subeV}
\vspace{-1ex}
m_{{\gamma^\prime }}\,\lesssim\,0.1\ {\rm eV}.
\vspace{-1ex}
\end{equation}
This minimal model can be supplemented by an axion-like spin-zero particle, coupled to the
minicharged fermions~\cite{Masso:2006gc}. A triangle diagram then leads to a coupling of the
axion-like particle to two photons.
The resulting production of axion-like particles gives an additional (to the one from
minicharged fermions)
contribution to the vacuum magnetic dichroism and birefringence.
One can then expect to have observable effects with even smaller values of
$\epsilon$, while
still not being in conflict with astrophysics.

The purpose of this letter is to point out that the required multiple $U(1)$ factors, the size of
kinetic-mixing, and suitable matter representations
to allow for detection in near future experiments occur very naturally within the
context of realistic extensions of the Standard Model (SM)
based on string theory.  It is a feature of our approach that we do \emph{not} construct a model
specifically for the purpose of producing minicharged particles
but instead argue that the required minicharged
particles are a generic, but also testable prediction of a large class of string theory models.

Let us begin by recalling the essentials of gauge kinetic-mixing.
It arises generally in theories that have, in addition to
some visible $U(1)_a$, at least one other additional $U(1)_b$ factor in
a hidden sector. In the basis in which the interaction terms have the canonical
form, the pure gauge part of the Lagrangian for an arbitrary
${U}(1)_a\times {U}(1)_b$
theory can be written as (generalization to more than one additional $U(1)$ factor
is straightforward)
\begin{equation}
{\mathcal L}_{\rm gauge} =
-\frac{1}{4} F^{\mu\nu}_{(a)} F_{(a)\mu\nu}
-\frac{1}{4} F^{\mu\nu}_{(b)} F_{(b)\mu\nu}
+\frac{\chi}{2} F^{\mu\nu}_{(a)} F_{(b)\mu\nu},
\end{equation}
where $\chi$ parametrizes the mixing. This parameter is directly linked to charge
shifts.
Notably, starting for example with two fermion species
$f_a$ and $f_b$ with charges $(e,0)$ and $(0,e)$, respectively, under
${U}(1)_a\times {U}(1)_b$, one finds that their
charges, after diagonalization of the gauge kinetic term, are shifted by the amount
\vspace{-1.5ex}
\begin{equation}
\epsilon \simeq \chi,
\vspace{-0.5ex}
\end{equation}
to leading order in $\chi\ll 1$~\cite{Holdom:1985ag}.
In other words, the hidden sector fermion $f_b$ picks up a small charge
$\epsilon \simeq \chi$ under the visible sector $U(1)_a$.

\begin{figure}
\begin{center}
\subfigure[]{
\includegraphics[width=3.5cm]{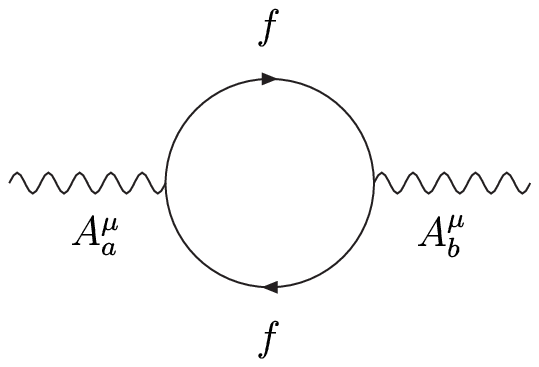}
\label{fig:QFT}}
\hspace{1cm}
\subfigure[]{
\scalebox{0.85}[0.85]{
\hspace{-0.9cm}
\includegraphics[width=3cm]{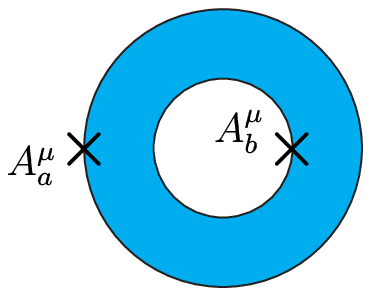}
}\label{fig:string}}
\end{center}
\vspace{-6ex}
\caption[...]{(a) One-loop diagram which contributes to kinetic-mixing in field theory,
and (b) its equivalent in open string theory (from Ref.~\cite{Abel:2004rp}).
\label{fig:loop}}
\end{figure}

In view of the current experimental sensitivity, the following
question immediately arises in this context:

{\it Are there theoretically appealing and phenomenologically viable
theories which naturally lead to a value of
$\chi\simeq \epsilon \sim 10^{-7}$
for the kinetic-mixing parameter?}

We will argue that the answer is yes.
Many SM extensions,  and indeed most extensions
coming from string theory, predict additional hidden $U(1)$ factors
which can give rise to the kinetic-mixing phenomenon
(e.g.,~\cite{Dienes:1996zr,Lust:2003ky,Abel:2003ue,Abel:2004rp,Batell:2005wa,Blumenhagen:2006ux}).

In the context of field theory, a non-zero value of $\chi$
can be induced quite generally at the one-loop level when there are states
which are simultaneously charged under both the visible $U(1)$ and the hidden
$U(1)$ factors~\cite{Holdom:1985ag}
(cf. the field theory diagram in Fig.~\ref{fig:QFT}).
The expected magnitude of $\chi$ in the field theory setting may be
estimated ~\cite{Holdom:1985ag,Dienes:1996zr}
by considering the contribution of two fermions with charges $(e_a,e_b)$ and
$(e_a,-e_b)$ and masses $m$ and $m^\prime$, respectively.
Their joint contribution to $\chi$ is given by~\cite{Holdom:1985ag}
\begin{equation}
\label{field-theory}
\chi \simeq \left( e_ae_b/(6\pi^2)\right) \log\left( m^{\prime}/m\right)
,
\end{equation}
and can be of order  $10^{-7}$, for natural values of
$e_ae_b/(4\pi)\sim \alpha$, with $\alpha =e^2/(4\pi)$ being the electromagnetic fine-structure
constant, and
nearly degenerate masses, $m^\prime/m\sim 1.0002$.
Kinetic-mixing could be avoided in such theories if the particle spectrum
has some particular  properties, as
discussed in Ref.~\cite{Dienes:1996zr}. For example, if one or
both of the $U(1)$ gauge factors sits within an unbroken non-abelian gauge symmetry,
then kinetic-mixing is not possible simply because of the tracelessness of the generators.

In the string theory context (e.g.,~\cite{Polchinsky:book}),
the story is more subtle and more varied,
but quite generally there are clear hints that
string theory naturally leads to a generation of
$\chi\neq 0$~\cite{Dienes:1996zr,Lust:2003ky,Abel:2003ue,Abel:2004rp,Blumenhagen:2006ux},
 and can naturally
give values of the right order of magnitude. The expected size of $\chi$ in a
particular string theory setting can be investigated by performing
the equivalent one-loop string theory calculation (cf. Fig.~\ref{fig:string}).

Weakly coupled heterotic closed string models (in which the string scale,
$M_s\approx 5 \times 10^{17}$~GeV, is close to the Planck scale, $M_P=1.2\times 10^{19}$~GeV)
were treated this way in Ref.~\cite{Dienes:1996zr}. In these models, even if
the particular string vacuum does not have the $U(1)$ explicitly embedded
in a non-abelian group,
there remains a ``memory'' of the underlying non-abelian structure,
and the contributions vanish at leading order.
Nevertheless, various effects below the string
scale reintroduce kinetic-mixing. In particular, low energy supersymmetry (SUSY)
breaking will {\em always} split the matter multiplets contributing to the
kinetic-mixing by an amount
of order the supersymmetry breaking scale in the hidden sector.
If supersymmetry breaking is mediated by gravity, then the latter
is roughly $M_{\rm SUSY}\sim 10^{11}$~GeV and
 $\chi\sim 10^{-7}$ can naturally be obtained~\cite{Dienes:1996zr}.

\begin{figure}
\begin{center}
\includegraphics[%
  bb=100bp 170bp 499bp 599bp,
  scale=0.25]{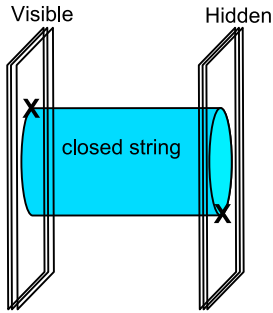}
\end{center}
\caption[...]{Kinetic-mixing in open string models with SUSY
breaking on ``hidden'' branes. The
visible sector consists of a phenomenologically well determined
supersymmetric configuration
of D3-branes at a fixed point in the 6 dimensional
compact manifold, possibly with D7-branes
passing through to cancel local tadpoles.
Global absence of tadpoles is assumed to require additional
branes and/or anti-branes in the bulk. Closed string
interactions are mediated from hidden to visible sector by
cylinder diagrams, and are equivalent to Fig.~\ref{fig:string}.
\label{fig:cylinder}}
\end{figure}

Other interesting classes of string models are
mostly based on configurations of Dirichlet (D)-branes and/or fluxes.
(D$p$ branes are membrane-like objects with $p$ spatial dimensions,
where open strings can be attached,
thereby becoming matter and gauge fields.)
Particularly interesting are the intermediate
scale models with $M_s\sim \sqrt{M_W M_P}\sim 10^{11}$~GeV. The
so-called
bottom-up approach~\cite{Aldazabal:2000sa,Blumenhagen:2000wh}
 is a good way
to ascertain general properties of these models;
the idea is to fix the gauge properties
of the SM or minimal supersymmetric SM (MSSM) {\em locally} in the compactified space using D-branes,
without fixing the {\em global} details of the compactification. The
latter have only a minor affect on the visible sector phenomenology
and may vary. However, the {\em global} set-up
almost certainly requires anti-D-branes in the bulk which absorb
certain types of  unwanted  tadpoles  but leave supersymmetry
broken.
Inevitably, these set-ups predict kinetic-mixing and therefore the existence of
minicharged particles~\cite{Abel:2003ue,Abel:2004rp}. This phenomenon generally
arises in these models because the hidden sector of $\overline {\rm D}$-branes
in the bulk carry additional hidden $U(1)$ factors (possibly emerging from $U(N)$),
which interact with the visible sector MSSM
branes by exchanging closed string modes through the bulk (cf. Fig.~\ref{fig:cylinder}).
Such a closed string exchange (a cylinder diagram)
can also be interpreted in the ``open string channel'' as a
kinetic-mixing diagram as shown in Fig.~\ref{fig:string};
the one-loop open string diagram has a heavy string in the loop
stretched between brane and anti-brane.
The masses of the modes in the loop are given by their
stretching energy proportional to their length
(i.e. the distance between brane and anti-brane). The reason, the
one-loop contributions do not cancel, is that the presence of anti-branes breaks
supersymmetry, which is in fact integral to this particular scenario.
Consequently there is a residual contribution
to kinetic-mixing and hence $\chi$, which is again given by the
amount of supersymmetry breaking.

\begin{figure}
\begin{center}
\includegraphics[bbllx=49,bblly=224,bburx=580,bbury=607,width=6.5cm]{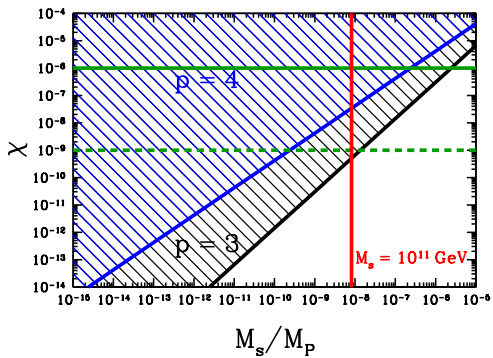}
\end{center}
\vspace{-5.5ex}
\caption[...]{Possible values for the kinetic-mixing parameter $\chi$, as a
function of the string scale, in the bottom-up approach discussed in the text.
Values in the shaded region above the black (lower) and blue (upper) lines
are predicted in models with D3- and D4-branes, respectively, for
$R/r\geq 1$. The red (vertical) line gives the largest string scale allowed
by phenomenology in these models. The area above the green solid (horizontal) line
is excluded by current experiments
searching for minicharged particles
while the green dashed line gives an idea of the expected sensitivity in the near future.
\label{fig:parameters}}
\end{figure}

Consider non-degenerate radii, with our three infinite dimensions,
$d+p-3$ large dimensions of radius $R_{i=4..p+d}=R$, and $9-d-p$ small space dimensions of radius
$R_{i=p+d+1..10}=r$, with the visible and hidden sectors
living on stacks of D$p$-branes wrapping the small dimensions
(where $p\geq 3$), and with the distance between
hidden and visible branes being generically of size the
compact dimension. The result for the mixing parameter
when the hidden $U(1)_b$ is unbroken can be written as~\cite{Abel:2003ue}
\vspace{-4ex}
\begin{eqnarray}
\!\!\chi &\sim & \pi\frac{\scriptsize \alpha_p }{N}
X_{a}X_{b}
\left( \frac{2^{(8-p)/2}}{\alpha_p}\frac{M_s}{M_P}\right)^{\frac{2(5-p)}{6-p}}
\left( \frac{R}{r} \right)^{\frac{d-p+3}{6-p}} .
\label{chi_intermediat_degen2}
\end{eqnarray}
(For the present discussion the spatial dimensions of the
branes is assumed to be $p$ for both visible and hidden sectors; for
the more general set-up, see Ref.~\cite{Abel:2003ue}.)
The integer $N$ is a factor corresponding to the  $Z_N$ point-group symmetry of
the fixed point, a typical value being $N=3$. $X_{a,b}$
represent factors coming from the traces of Chan-Paton matrices in
the vertex operators for photon emission off the open string loop, corresponding to the
crosses in  Figs.~\ref{fig:string}, \ref{fig:cylinder};
typically $X_{a,b}\sim 1$, but we will comment more precisely on the
meaning of these factors below.
Finally, $\alpha_p$ is the value of the gauge coupling, for which one may take
$\alpha_p\sim 1/24$ (the MSSM unification value).
For example if $p=d=3$, then
\vspace{-2ex}
\begin{equation}
\chi  \simeq 5\times 10^{-10}
\,(R/r)\, \left(M_{s}/10^{11}\rm{GeV}\right)^{\frac{4}{3}}
,\vspace{-1ex}
\end{equation}
so that an $R/r\sim 10^4$ would give the right value.
Varying $p\geq 3$ and $d$, there is considerable freedom to reduce the ratio $R/r$, whilst
still having $\chi\sim 10^{-7}$ (cf. Fig.~\ref{fig:parameters}).
 (Note that the strong bounds on $M_s$ obtained in \cite{Abel:2003ue}
from the astrophysical bounds
on $\epsilon$ do not apply in general due to the charge screening mechanism
proposed in \cite{Masso:2006gc}.)


Before continuing with the generic consideration of volume and scale
dependence, we should briefly digress, to consider a subtle and
somewhat technical issue, namely the question of whether such
$U(1)$s can remain massless (or at least very much lighter than the
string scale) but still kinetically mix\footnote{We thank Mark
Goodsell for extensive input and collaboration on these and related
issues which are discussed in detail in Ref.~\cite{thefuture}.}.
 This is a delicate question because the kinetic-mixing
diagram is also the diagram for a mass term mixing visible and hidden photons,
and the one-loop open string diagrams one evaluates in fact correspond to two terms in the
Lagrangian of the form
\begin{equation}
\label{mass}
 m_{ab}^2 A^a_{\mu} A_b^{\mu} + \chi_{ab} F^{(a)}_{\mu\nu} F^{(b)\, \mu\nu}\, .
\end{equation}
The mass term $m_{ab}$ is a St\"uckelberg mass mixing, which is
associated with mixed anomalies and their cancellation via the Green-Schwarz
mechanism~\cite{Antoniadis:2002cs,Anastasopoulos:2003aj}.
Anomaly free $U(1)$s must have $m_{ab}=0$. (Also note that
$U(1)$s that are anomaly-free in 4d may still get St\"uckelberg masses
due to 6d anomalies.)  In the second term, $\chi_{ab}$ is the kinetic
mixing parameter.  Since both of the terms in Eq.~\eqref{mass} arise
from the same diagram, how can $\chi_{ab}$ be non-vanishing in an
anomaly free theory where $m_{ab}=0$?

The answer is that in order to
get a contribution to the St\"uckelberg mass one has to extract a
$1/k^2$ pole from the appropriate one-loop integral.  From the closed
string point of view this corresponds to the St\"uckelberg mass only
getting contributions from {\em massless} closed string modes.  Such
contributions are blind to the location in the compact dimensions of
the different sources.  The non-pole contributions in this integral
gives rise to $\chi_{ab}$.  Importantly these contributions to
$\chi_{ab}$ are from both massless {\em and} massive Kaluza Klein
modes.  The latter certainly do care about the location of the sources
in the compact dimensions, and so contributions to $\chi_{ab}$ do not
generally cancel even though the contributions to $m_{ab}$ must.
Therefore $\chi$ arises for $U(1)$s that are anomaly free provided
that the anomaly-free combination is from branes that are located at
different points or wrapping different cycles.

This is in fact generic in large volume compactifications.
For example for branes parallel to orientifold planes, the
anomaly-free $U(1)$ comes from the original brane plus its
displaced orientifold image, and the two contributions to
kinetic mixing do not cancel.
In a future publication~\cite{thefuture}, this will be confirmed by
showing explicit constructions
where kinetic mixing occurs between
anomaly-free $U(1)$s even if they come purely from D-branes
in completely supersymmetric and tadpole-free
configurations.


Let us now return to the generic implications for
the volume and scale dependence that can be derived from Eq. (\ref{chi_intermediat_degen2}),
and comment on the fermion sector.
A crucial ingredient for the electric charge screening mechanism in
the stellar plasma is that there have to be {\em two} hidden sector
$U(1)$s, and the hidden sector fermions have to have charge
$(0,e,-e)$ under the visible and the two hidden sector $U(1)$
factors, respectively~\cite{Masso:2006gc}.
Note that this is generic in open string models
with hidden D-brane sectors, since hidden sector
fields arising from open strings stretched between hidden sector branes
naturally fall into the bifundamental represention
of the two hidden sector $U(1)$'s.

\begin{figure}
\begin{center}
\includegraphics[%
  bb=200bp 520bp 400bp 599bp,
  scale=0.58]{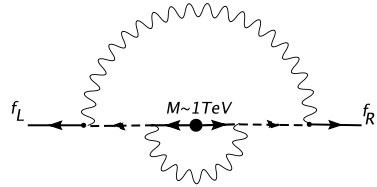}
\end{center}
\vspace{-1.2ex}
\caption[...]{Two-loop induced fermion masses in the hidden sector
are directly related to the supersymmetry breaking. The
dashed lines and gauge bosons are all stretched between visible and hidden
branes, while the internal propagator is a fermion mass propagator of order
1~TeV.
\label{fig:see-saw}}
\end{figure}

The remaining question is
whether there is any reason to expect the mass of the
minicharged particles to be $\sim 0.1$ eV or smaller, and indeed
there is. Since one of the $U(1)$s is by assumption unbroken,
it is natural to expect some fermions on the hidden brane to
be initially massless. However, as we have seen, the $U(1)_b$ mixes with the
visible sector symmetries, and of course here the MSSM requires
mass terms, namely the $\mu$-term for the Higgs. Generally,
these induce two-loop mass terms in the hidden sector, as shown
in Fig.~\ref{fig:see-saw}. These contributions are
diluted by the same volume factors ($V_{||},V_\perp$)
that cause the dilution of gravity, given by~\cite{Ibanez:1998rf}
\begin{equation}
M_s^2/M_P^2\sim  \alpha_p^{2}\, V_{||}/V_\perp\, ;
\end{equation}
at one-loop the stretched states get a mass-splitting
(the inner loop of Fig.~\ref{fig:see-saw}) of order
$
({V_{||}}/{V_\perp}) {M_s^2}/{\mu} \, ,
$
where $\mu \sim 1$~TeV, and at two-loops
the diagram receives another volume
factor $V_{||}/{V_\perp}$. In total, therefore,
the mass induced in the hidden sector is
\begin{equation}
m_{\rm hidden} =   \alpha_p^{-4}
\,(M_s^6/M_P^4\, \mu )\,
\sim \alpha_p^{-4}\,(M_W^2/M_P) \,,
\end{equation}
which is roughly of the right order of magnitude for $\alpha_p\sim 1/24$.
Note, that no new scales
beyond what is assumed for the MSSM have been introduced. The conclusion
is quite general: sub~eV masses are induced for hidden sector fermions if
there are fermions with mass $\sim 1$~TeV
in the visible sector.

Therefore, both closed and open string models not only
predict the necessary extra
$U(1)$ factors and correct fermion representation,
but can also accommodate values of the
kinetic-mixing parameter and fermion masses that may allow for detection in the near future.

As far as the gauge boson mass is concerned, the good news is that,
naturally, a Higgs appears in the bifundamental representation of the hidden $U(1)$s,
leaving automatically one mixed $U(1)$ massless, as required in the charge screening
mechanism of~\cite{Masso:2006gc}. It remains to be seen whether one can
come up with a mechanism, perhaps based on accidental symmetries, to stabilize its small sub-eV scale
(cf. Eq.~(\ref{subeV})). Finally, there may still be room for
an additional light spin-zero particle coupled to the hidden sector fermions
which could then play the role of an axion-like particle~\cite{Masso:2006gc,Jaeckel:2006id}.

There are a number of exciting possibilities to test such a scenario in
laboratory experiments, allowing for experimental insights into string
theory with less model dependence than astrophysical or cosmological considerations.

The existence of minicharged particles can be tested~\cite{Gies:2006ca} by improving the
sensitivity of instruments for the detection
of vacuum magnetic birefringence and
dichroism~\cite{Cameron:1993mr,Zavattini:2005tm,Chen:2003tp,Rizzo:Patras,Pugnat:2005nk,Heinzl:2006xc,Chen:2006cd}.
Another sensitive tool is Schwinger pair production
in strong electric fields, as they exist, for example, in accelerator
cavities~\cite{Gies:2006hv}. A classical probe is the search for invisible orthopositronium
decays~\cite{Rubbia:2004ix,Badertscher:2006fm}. We expect that all these laboratory experiments will
probe into the range $\epsilon\sim 10^{-9}-10^{-6}$.
Hidden-sector $U(1)$ gauge bosons \cite{Ahlers:2007qf} and additional axion-like
particles~\cite{Gasperini:1987da,VanBibber:1987rq}, coupled to the minicharged fermions, may be
observed in photon regeneration
experiments~\cite{Ringwald:2003ns,Pugnat:2005nk,Rabadan:2005dm,Cantatore:Patras,Kotz:2006bw,Baker:Patras,Rizzo:Patras,Ehret:2007cm,Afanasev:2006cv,OSQAR,PVLASLSW}
some of which have recently published data~\cite{Robilliard:2007bq,Chou:2007zz}. These searches are complementary to
and presently more sensitive than collider techniques based on
the effect of kinetic-mixing on precision electroweak
observables~\cite{Babu:1997st,Kumar:2006gm,Feldman:2007wj}.

{\it In conclusion,} \JJ{many string theory models
with intermediate string scales $M_s\sim 10^{11}$~GeV
and/or large volumes
predict the existence of minicharged particles with \JJ{$\epsilon\gtrsim  10^{-9}$}, testable with
near future laboratory experiments.}

\vspace{-2.5ex}

\bigskip
\bigskip

\centerline{\bf  Acknowledgements}

\bigskip

\noindent We are grateful to Mark Goodsell for informative and helpful discussions.

\end{document}